\documentclass[twoside]{article}
\usepackage{amssymb,amsmath}
\usepackage{qic,epsfig}

\newcommand{\cA}{\mathcal{A}}
\newcommand{\cB}{\mathcal{B}}
\newcommand{\cC}{\mathcal{C}}

\newcommand{\cP}{\mathcal{P}}

\newcommand{\cS}{\mathcal{S}}

\newcommand{\Id}{\mathbf{I}}
\newcommand{\tr}{\text{Tr}}

\newtheorem{conjecture}{Conjecture}

\renewcommand{\thefootnote}{\fnsymbol{footnote}}  

\begin{document}
\setlength{\textheight}{8.0truein}    

\runninghead{Unextendible Mutually Unbiased Bases from Pauli Classes }
            {P.Mandayam, S.Bandyopadhyay, M.Grassl, and W. K.Wootters }

\normalsize\textlineskip
\thispagestyle{empty}
\setcounter{page}{1}


\vspace*{0.88truein}

\alphfootnote

\fpage{1}

\centerline{\bf
UNEXTENDIBLE MUTUALLY UNBIASED BASES FROM PAULI CLASSES}
\vspace*{0.37truein}
\centerline{\footnotesize
PRABHA MANDAYAM}
\vspace*{0.015truein}
\centerline{\footnotesize\it The Institute of Mathematical Sciences, Taramani}
\baselineskip=10pt
\centerline{\footnotesize\it Chennai
  - 600113, India.}
\vspace*{10pt}
\centerline{\footnotesize
SOMSHUBHRO BANDYOPADHYAY}
\vspace*{0.015truein}
\centerline{\footnotesize\it Department of Physics and Center for Astroparticle
  Physics and Space Science, Bose Institute, Bidhan Nagar}
\baselineskip=10pt
\centerline{\footnotesize\it Kolkata - 700091, India.}
\vspace*{10pt}
\centerline{\footnotesize
MARKUS GRASSL}
\vspace*{0.015truein}
\centerline{\footnotesize\it Centre for Quantum Technologies, National University of
  Singapore}
\baselineskip=10pt
\centerline{\footnotesize\it Singapore 117543, Singapore.}
\vspace*{10pt}
\centerline{\footnotesize
WILLIAM K. WOOTTERS}
\vspace*{0.015truein}
\centerline{\footnotesize\it Department of Physics, Williams College, Williamstown}
\baselineskip=10pt
\centerline{\footnotesize\it MA - 01267, USA.}
\vspace*{0.225truein}

\vspace*{0.21truein}


\abstracts{We provide a construction of sets of $d/2+1$ mutually unbiased bases (MUBs) in dimensions $d=4,8$ using maximal commuting classes of Pauli operators. We show that these incomplete sets cannot be extended further using the operators of the Pauli group. However, specific examples of sets of MUBs obtained using our construction are shown to be {\it strongly unextendible}; that is, there does not exist another vector that is unbiased with respect to the elements in the set. We conjecture the existence of such unextendible sets in higher dimensions $d=2^{n} (n>3) $ as well.} {Furthermore, we note an interesting connection between these unextendible sets and state-independent proofs of the Kochen-Specker Theorem for two-qubit systems. Our construction also leads to a proof of the tightness of a $H_{2}$ entropic uncertainty relation for any set of three MUBs constructed from Pauli classes in $d=4$. }{}

\vspace*{10pt}

\vspace*{3pt}

\vspace*{1pt}\textlineskip    
\section{Introduction}
Two orthonormal bases $\cA = \{|a_{i}\rangle, i=1,\ldots, d\}$ and
$\cB = \{|b_{j}\rangle, j=1, \ldots, d\}$ of the $d$-dimensional
Hilbert space $\mathbb{C}^{d}$ are said to be {\it mutually unbiased}
if for every pair of basis vectors $|a_{i}\rangle \in \cA$ and
$|b_{j}\rangle\in \cB$,
\begin{equation}
\vert\langle a_{i}|b_{j}\rangle\vert = \frac{1}{\sqrt{d}}, \forall i,
j = 1,\ldots, d.
\end{equation}
Two bases $\cA$ and $\cB$ that are mutually unbiased have the property
that if a physical system is prepared in an eigenstate of basis $\cA$
and measured in basis $\cB$, all outcomes are equally probable. A set
of orthonormal bases $\{\cB_{1}, \cB_{2}, \ldots, \cB_{m}\}$ in
$\mathbb{C}^{d}$ is called a set of mutually unbiased bases (MUBs) if
every pair of bases in the set is mutually unbiased. MUBs play an
important role in our understanding of complementarity in quantum
mechanics and are central to several key quantum information processing tasks including quantum cryptography and state tomography.

MUBs form a minimal and optimal set of orthogonal measurements for
quantum state tomography~\cite{Ivanovic81,WF89}. To specify a general
density matrix $\rho \in \mathbb{C}^{d}$, which is Hermitian and has
$\tr(\rho)=1$, one needs $d^{2}-1$ real parameters. Since measurements
within a particular basis set can yield only $d-1$ independent
probabilities, one needs $d+1$ distinct basis sets to provide the
required total number of $d^{2}-1$ independent
probabilities. Correspondingly, the maximal number of MUBs that can
exist in $d$-dimensional Hilbert space is $d+1$ and explicit
constructions of such maximal sets are known when $d$ is a prime
power~\cite{WF89, BBRV02, LBZ02}. However, in composite dimensions
while smaller sets of MUBs have been constructed~\cite{Grassl04,WB05},
the question as to whether a complete set of MUBs exists in
non-prime-power dimensions still remains unresolved. We refer to~\cite{mubreview_2010} for a recent review of the various constructions and applications of MUBs.

MUBs also play an important role in quantum cryptographic protocols,
since they correspond to measurement bases that are most
`incompatible', as quantified by uncertainty relations. A set of
measurement bases is said to be {\it maximally incompatible} if they
satisfy a maximally strong uncertainty relation. Being mutually
unbiased is a necessary condition for a set of measurement bases to be
maximally incompatible~\cite{WWsurvey}. The security of cryptographic
tasks such as quantum key distribution~\cite{bb84} and two-party
protocols using the noisy-storage model~\cite{KWW12} relies on this
property of MUBs. In particular, protocols based on higher-dimensional
quantum systems with larger numbers of unbiased basis sets can have
certain advantages over those based on
qubits~\cite{dlevel_qkd,MW11}. It is therefore important for
cryptographic applications to find sets of MUBs that satisfy strong
uncertainty relations.

Related to the question of finding complete sets of MUBs is the
important concept of {\it unextendible} MUBs. A set of MUBs
$\{\cB_{1}, \cB_{2}, \ldots,\cB_{m}\}$ in $\mathbb{C}^{d}$ is said to
be {\it unextendible} if there does not exist another basis in
$\mathbb{C}^{d}$ that is unbiased with respect to all the bases
$\cB_{j}, j=1,\ldots,m$. In this paper we show the existence of unextendible sets of MUBs even in systems for which a complete set of MUBs is known to exist.


We follow a standard construction of MUBs based on finding mutually
disjoint, maximal commuting classes of tensor products of Pauli
operators~\cite{BBRV02, LBZ02}. It is always possible to find a
partitioning of the $n$-qubit Pauli operators in $d=2^{n}$ into $d+1$
disjoint maximal commuting classes, the common eigenbases of which
form a complete set of $d+1$ MUBs~\cite{BBRV02, LBZ02}. Here, we show
that there exist smaller sets of $k<d+1$ commuting classes $\{\cC_{1},
\cC_{2},\ldots, \cC_{k}\}$ in $d=4,8$ that are {\it unextendible} in
the following sense---no more maximal commuting classes can be formed
out of the remaining $n$-qubit Pauli operators that are not contained
in $\cC_{1}\cup\cC_{2}\ldots\cup\cC_{k}$. The eigenbases of
$\{\cC_{1},\ldots,\cC_{k}\}$ thus give rise to a set of $k$ MUBs which
cannot be extended using joint eigenvectors of maximal sets of
commuting Pauli operators.  We call such sets {\it weakly unextendible}.

We also obtain examples of {\it strongly unextendible} sets of MUBs
using our construction of unextendible classes in $d=4,8$, that is,
there does not exist even a single vector unbiased with respect to the
bases in these sets. For two-qubit systems, our construction of
unextendible sets of maximal commuting Pauli classes enables us to
prove the tightness of an entropic uncertainty relation. Furthermore,
we also demonstrate an interesting connection between unextendible
sets of classes and state-independent proofs of the Kochen-Specker
Theorem~\cite{KS67, mermin90}.

The rest of the paper is organized as follows. We formally define weak
and strong unextendibility in Section~\ref{sec:prelim} and review the
standard construction of MUBs from maximal Pauli classes. In
Section~\ref{sec:unext} we state our main results on constructing
unextendible Pauli classes in $d=4,8$, detailed proofs of which are
given in Appendix \ref{sec:proof_d4} and
\ref{sec:proof_d8}. In~Section~\ref{sec:strong_unext} we present
examples of sets of MUBs obtained using our construction that are in
fact strongly unextendible. Finally, we discuss properties and
potential applications of such unextendible sets in
Section~\ref{sec:props}.


\setcounter{footnote}{0}
\renewcommand{\thefootnote}{\alph{footnote}}

\section{Preliminaries}\label{sec:prelim}

\subsection{Construction of MUBs from Maximal Commuting Operator Classes}

Let $\cS$ be a set of $d^{2}$ mutually
orthogonal\footnote{Orthogonality is defined with respect to the
  Hilbert-Schmidt norm.  Unitary operators $U_{i}$ and $U_{j}$ are
  said to be orthogonal if $\tr[U_{i}^\dagger U_{j}] = 0$.}~ unitary
operators in $\mathbb{C}^{d}$. This set of $d^{2}$ operators
(including the identity operator $\mathbf{I}$) constitutes a basis for
the space of $d\times d$ complex matrices. A standard construction of
MUBs outlined in~\cite{BBRV02, LBZ02} relies on finding classes of
commuting operators, with each class containing $(d-1)$
mutually orthogonal commuting unitary matrices different from identity.

\begin{definition}[Maximal Commuting Operator Classes]
A set of subsets $\{\cC_{1} ,\cC_{2} ,\ldots, \cC_{L}$ $\mid
\cC_{j}\subset \cS\setminus\{\mathbf{I}\}\}$ of size $|\cC_{j}|=d-1$
constitutes a (partial) partitioning of $\cS\setminus\{\mathbf{I}\}$
into mutually disjoint maximal commuting classes if the subsets $\cC_{j}$ are
such that (a) the elements of $\cC_{j}$ commute for all $1\leq j\leq
L$ and (b) $\cC_{j}\cap\cC_{k}=\emptyset$ for all $j \neq k$.
\end{definition}

In the rest of the paper, we often use the term {\it operator classes}
to refer to such mutual disjoint maximal commuting classes. In
particular, we use the term {\it Pauli classes} to refer to mutual
disjoint maximal commuting classes formed out of the $n$-qubit Pauli
group $\cP_{n}$.

The correspondence between maximal commuting operator classes and MUBs
is stated in the following Lemma, originally proved in~\cite{BBRV02}.
\begin{lemma}\label{lem:correspondence}\
The common eigenbases of $L$ mutually disjoint maximal commuting operator
classes form a set of $L$ mutually unbiased bases.
\end{lemma}

\subsection{Unextendibility of MUBs and Operator Classes }

\begin{definition}[{\bf Unextendible Sets of MUBs}]\label{def:unext_general}
A set of MUBs  $\{\cB_{1}, \cB_{2}, \ldots, \cB_{L}\}$ is
unextendible if there does not exist another basis that is unbiased
with respect to the bases $\cB_{1}$, \ldots, $\cB_{L}$.
\end{definition}

For example, it is known that in dimension $d=6$, the eigenbases of $\hat{X}, \hat{Z}$ and $\hat{X}\hat{Z}$ are an unextendible set of three MUBs~\cite{Grassl04}, where $\hat{X}$ and
$\hat{Z}$ are the generators of the Weyl-Heisenberg group. This has the important consequence that starting with the Weyl-Heisenberg generators we cannot hope to obtain a complete set of seven MUBs in $d=6$. This result has been generalized recently by removing the restriction that the second basis be related to the Weyl-Heisenberg group. In particular, it is shown that allowing for different choices for the second basis, by thoroughly sampling the set of currently known complex Hadamard matrices in $\mathbb{C}^6$, also leads to unextendible sets of three MUBs in $d=6$~\cite{Brierley09}.

Furthermore, the existence of an infinite family of unextendible
triplets of MUBs in $d=6$ has been proved using the Fourier family of
Hadamard matrices~\cite{Jaming09}. Similarly, it has been shown that
starting with two product bases in $d=6$ also yields unextendible sets
of three MUBs~\cite{McNulty12}. These results further lend credence to
the long-standing conjecture~\cite{Zauner_thesis} that {\it any} set
of three MUBs in $d=6$ is in fact unextendible to a complete set.

Moving away from six dimensions, the set of three MUBs obtained in
$d=4$ using Mutually Orthogonal Latin Squares (MOLS) \cite{WB05} is an
example of an unextendible set of MUBs in prime-power
dimensions~\cite{BSTW05}.

A stronger notion of unextendibility can be defined as follows.
\begin{definition}[{\bf Strongly Unextendible Sets of MUBs}]
A set of MUBs $\{\cB_{1}, \cB_{2}, \ldots, \cB_{L}\}$ in
$\mathbb{C}^{d}$ is said to be strongly unextendible if there does not
exist any vector $v \in \mathbb{C}^{d}$ that is unbiased with respect
to the bases $\cB_{1}$, \ldots, $\cB_{L}$.
\end{definition}
The eigenbases of $\hat{X}, \hat{Z}$ and
$\hat{X}\hat{Z}$ in $d=6$ are known to be strongly
unextendible~\cite{Grassl04}. It is further conjectured that the set
of three MUBs obtained as eigenbases of $\hat{X}, \hat{Z}$ and
$\hat{X}\hat{Z}$ are strongly unextendible in any even dimension
($d=2m$), a conjecture that has been verified for $d\leq
12$~\cite{Grassl09}. More recently, it was shown that any set of three
mutually unbiased product bases in dimension six is strongly
unextendible~\cite{McNulty12_ijqi}.

The correspondence between MUBs and maximal commuting operator classes
gives rise to a weaker notion of unextendibility, based on
unextendible sets of such classes.
\begin{definition}[{\bf Unextendible Sets of Operator Classes}]\label{def:unext_class}
A set of mutually disjoint maximal commuting classes $\{\cC_{1},
\cC_{2}, \ldots, \cC_{L}\}$ of operators drawn from a unitary basis $\cS$ is said to
be unextendible if no other maximal class can be formed out of the
remaining operators in $\cS\setminus(\{\mathbf{I}\}\cup\bigcup_{i=1}^{L}\cC_{i})$.
\end{definition}
The eigenbases of such an unextendible set of classes form a weakly unextendible set of MUBs, as defined below.
\begin{definition}[{\bf Weakly Unextendible Sets of MUBs}]\label{def:unext}
Given a set of MUBs $\{\cB_{1}, \cB_{2}, \ldots,$ $\cB_{L}\}$ that are
realized as common eigenbases of a set of $L$ operator classes
comprising operators from $\cS$, the set $\{\cB_{1}, \cB_{2}, \ldots,
\cB_{L}\}$ is weakly unextendible if there does not exist another
unbiased basis that can be realized as the common eigenbasis of a
maximal commuting class of operators in $\cS$.
\end{definition}

For example, consider the following three Pauli classes in $d=4$:
\begin{alignat}{5}
\cC_{1} &{}= \{Y\otimes Y, I\otimes Y, Y\otimes I \}, \nonumber \\
\cC_{2} &{}= \{ Y\otimes Z, Z\otimes X, X\otimes Y \}, \nonumber \\
\cC_{3} &{}= \{ X\otimes I, I\otimes Z, X\otimes Z \} \label{eq:pauli3}
\end{alignat}
Here, $X,Y,Z$ denote the standard single-qubit Pauli operators. The
partitioning above makes use of only nine of the fifteen two-qubit
Pauli operators that constitute the set
$\cP_{2}\setminus\{\mathbf{I}\}$ in $d=4$. It is easy to check by hand
that it is not possible to find one more set of three commuting
operators from the remaining set
\begin{equation}
 \{I\otimes X, X\otimes X, Y\otimes X, Z\otimes I, Z\otimes Y, Z\otimes Z\} \nonumber
\end{equation}
of six Pauli operators. The set of three classes in~\eqref{eq:pauli3} is thus unextendible, and their common eigenbases are therefore a set of  weakly unextendible MUBs. We note that this set of MUBs was obtained earlier~\cite{MWB10} via a construction of smaller sets of MUBs in dimension $d=2^{n}$ using the generators of the Clifford algebra. This set was observed to have interesting properties, in particular, saturating an entropic uncertainty relation (EUR) for the $H_{2}$ entropy. Here, we explicitly prove the tightness of the EUR using our construction of unextendible classes (see Section~\ref{sec:eur}).


\section{Construction of Unextendible Pauli Classes in $d=4,8$}\label{sec:unext}

In $d=2^{n}$ dimensions, the set $\cP_{n}\setminus\{\mathbf{I}\}$ of
all tensor products of Pauli matrices contains a total of $(4^{n}-1)$
operators.  As observed earlier, there exists a partitioning of these
$(d^{2}-1)$ operators in $\cP_{n}\setminus\{\mathbf{I}\}$ into a
complete set of $(d+1)$ mutually disjoint maximal commuting classes,
with each class containing $(d-1)$ $n$-qubit Pauli operators. We begin
by observing a few properties of such complete sets of Pauli classes
which provide some intuition into our construction of unextendible
Pauli classes.

\begin{itemize}
\item[(P1)] Each operator in $\cP_{n}\setminus\{\mathbf{I}\}$ commutes with
  $(\frac{4^{n}}{2} - 2)$ distinct operators, excluding itself and the
  identity operator.

\item[(P2)] Each maximal commuting class is in fact an Abelian group
  generated by a set of $n$ commuting operators. The remaining
  operators are simply products of these $n$ {\it generators}. For
  example, in $d=4$, each maximal commuting class is generated by two
  commuting Pauli operators. The third element of the class is simply the
  product of the two generators. Similarly, in $d=8$, every maximal
  commuting class is generated by three commuting operators, say,
  $U_{1}$, $U_{2}$, $U_{3}$. Then, the non-trivial elements in the
  class are given by:
\begin{equation}
\{U_{1}, U_{2}, U_{3}, U_{1}U_{2}, U_{1}U_{3}, U_{2}U_{3}, U_{1}U_{2}U_{3}\} \nonumber
\end{equation}

\item[(P3)] Given any two maximal commuting classes, the remaining
  $d-1$ maximal commuting classes that constitute a complete set can
  be realized as products of the operators in these two classes. That
  is, given the $d-1$ operators of $\cC_{i}$ and the $d-1$ operators
  in $ \cC_{j}$ ($i,j \in [1,d+1]$), the remaining $d-1$ classes can
  be obtained as products of these $(d-1)^{2}$ operators. This fact
  follows from the following Lemma.
\begin{lemma}\label{lem:generators2}
The in total $2n$ generators of any two disjoint maximal commuting classes
are {\it independent} of each other.
\end{lemma}
{\bf Proof:} Suppose that the $n$ generators of a class $\cC_{i}$ and the $n$
generators of $\cC_{j}$ $(j \neq i)$, are not independent of each
other.  This implies that at least one of the generators, say
$\sigma^{(j)}_1\in\cC_{j}$, can be expressed as a product of some
generators $\sigma^{(i)}_\mu\in\cC_{i}$ and some of the remaining
$n-1$ generators of $\cC_{j}$, that is, $\sigma^{(j)}_1 =
\sigma^{(i)}_1\dots\sigma^{(i)}_m\sigma^{(j)}_2\dots\sigma^{(j)}_\ell$. But
this would mean that the non-trivial operator
$\sigma^{(j)}_1\sigma^{(j)}_2\dots\sigma^{(j)}_\ell=\sigma^{(i)}_1\dots\sigma^{(i)}_m$
belongs to both $\cC_{i}$ and $\cC_{j}$, which are in fact disjoint
sets. Thus the generators of any two classes must be independent
of each other. Note, however, that if we include a third class
$\cC_{k}$, the generators of $\cC_{k}$ can be obtained as products of
the generators of $\cC_{i}$ and $\cC_{j}$. \square

\item[(P4)] Every operator in a given class $\cC_{i}$ commutes with
  at most $n-1$ generators of any other class $\cC_{j}$ $(j\neq i)$,
  and a total of $2^{n-1}-1$ operators in the class $\cC_{j}$. If an
  operator of $\cC_{i}$ were to commute with all $n$ generators of
  $\cC_{j}$, it would give rise to a set of $n+1$ independent
  commuting operators, leading to a total of $2^{n+1}-1$ commuting
  operators.  Such a set cannot exist in $d=2^{n}$ since the
  cardinality of a maximal commuting set is $2^{n}-1$.

\item[(P5)] No two elements of $\cC_{i}$ can commute with the same set
  of $2^{n-1}-1$ operators in a different class $\cC_{j}$ $(j \neq
  i)$. This is formally stated and proven in the following Lemma.
\begin{lemma}\label{lem:(n-1)commute}
Every operator $U_{i}\in\cC_{i}$ commutes with exactly $2^{n-1}-1$
elements in any other class $\cC_{j}$ $(j\neq i)$.  This set of
$2^{n-1}-1$ operators is unique, that is, no two elements of $\cC_{i}$
can commute with the same set of $2^{n-1}-1$ operators in a different
class $\cC_{j}$ $(j \neq i)$.
\end{lemma}
{\bf Proof:}
Suppose two operators $U_{i}, V_{i} \in \cC_{i}$ commute with the same
set of $2^{n-1}-1$ operators in $\cC_{j}$ $(j\neq i)$. This would
imply that $U_{i}, V_{i}$ commute with the same $n-1$ generators of
$\cC_{j}$, thus leading to a total of $n+1$ {\it independent}
operators that all commute. These $n+1$ operators will then generate a
set of $2^{n+1} -1$ commuting operators, but such a set cannot exist
in dimension $d=2^{n}$. Thus, no two operators of a class $\cC_{i}$ can
commute with the same set of $2^{n-1}-1$ operators in a different
class $\cC_{j}$ $(j\neq i)$. \square
\end{itemize}

\subsection{Weakly Unextendible Sets of Three MUBs in $d=4$}\label{sec:unext_d4}

We now state our central result on constructing weakly unextendible
MUBs in $d=4$, and give the proof in Appendix~\ref{sec:proof_d4}.
\begin{theorem}\label{thm:3max}
Given three Pauli classes $\cS_{1}$, $\cS_{2}$, $\cS_{3}$ that belong
to a complete set of classes in $d=4$, there exists exactly one more
maximal commuting class of Pauli operators $\cS$ (distinct from
$\cS_{1}$, $\cS_{2}$, $\cS_{3}$) that can be formed using the
operators in $\cS_{1}\cup\cS_{2}\cup\cS_{3}$.

The class $\cS$ along with the remaining two classes $\cS_{4}$ and
$\cS_{5}$ (in the complete set) form an unextendible set of Pauli
classes, whose common eigenbases form a weakly unextendible set of
three MUBs.
\end{theorem}

For example, consider a complete partitioning of the two-qubit Pauli
operators in $\cP_{2}\setminus\{\mathbf{I}\}$ as follows:
\begin{alignat}{5}
\cS_{1} &= \{Z\otimes I, I\otimes Z, Z\otimes Z \}\nonumber \\
\cS_{2} &= \{X\otimes I, I\otimes X, X\otimes X\} \nonumber \\
\cS_{3} &= \{X\otimes Z, Z\otimes Y, Y\otimes X\} \nonumber \\
\cS_{4} &= \{Y\otimes I, I\otimes Y, Y\otimes Y\} \nonumber \\
\cS_{5} &= \{Y\otimes Z, Z\otimes X, X\otimes Y\} . \label{eq:d4complete}
\end{alignat}
The unextendible set in~\eqref{eq:pauli3} is then constructed as
follows: $\cC_{3}$ is the unique Pauli class that can be formed using
the operators in $\cS_{1}, \cS_{2}, \cS_{3}$, whereas $\cC_{1}$ and
$\cC_{2}$ are simply the remaining two classes $\cS_{4}$ and
$\cS_{5}$.

Theorem~\ref{thm:3max} not only proves the existence of unextendible
sets of Pauli classes in $d=4$, but provides a way to construct
unextendible sets starting from any two Pauli classes.
\begin{corollary}\label{corr: unext3d4}
Given any two disjoint Pauli classes $\cC_{1}$ and $\cC_{2}$ in
$d=4$, there always exists a third class $\cC'_{3}$, of commuting
operators such that the common eigenbases of $\cC_{1}$, $\cC_{2}$,
$\cC'_{3}$ constitute a weakly unextendible set of three MUBs in
$d=4$.
\end{corollary}
{\bf Proof:} To see this, we first note that any two disjoint Pauli
classes $\cC_{1}$ and $\cC_{2}$ in $d=4$ can be extended to a complete
set of maximal commuting classes. Let,
\begin{equation*}
\cC_{1} = \{U_{1}, V_{1}, U_{1}V_{1}\}\quad\text{and}\quad
\cC_{2} = \{U_{2}, V_{2}, U_{2}V_{2}\}.
\end{equation*}
Each element in $\cC_{1}$ commutes with exactly one element in
$\cC_{2}$.  Assume without loss of generality that
$[U_1,U_2]=[V_1,V_2]=[U_1V_1,U_2V_2]=0$.
The remaining three classes are then given by
\begin{alignat*}{5}
\cC_{3} &= \{U_{1}V_{2}, U_{2}V_{1}, U_{1}V_{1}U_{2}V_{2}\},\\
\cC_{4} &= \{U_{1}U_{2}, V_{1}U_{2}V_{2}, U_{1}V_{1}V_{2}\},\\
\cC_{5} &= \{U_{1}V_{1}U_{2},U_{1}U_{2}V_{2},V_{1}V_{2}\}.
\end{alignat*}
Commutativity within each class can be shown by direct calculation.

Once we have the remaining three classes that form a complete set,
Theorem~\ref{thm:3max} guarantees that we can form exactly one more
maximal commuting Pauli class using the remaining three classes. Thus,
if we construct $\cC'_{3}$ following Theorem~\ref{thm:3max}, by
picking one element each from each of these classes, we end up with
three classes whose common eigenbases are a weakly unextendible set of
three MUBs in $d=4$. \square

Finally, we show that using the two-qubit Pauli operators, we cannot
find any unextendible sets of four MUBs in $d=4$.
\begin{theorem}\label{thm:d_extendible}
Given two Pauli classes $\cC_{1}$ and $\cC_{2}$ that belong to a
complete set of maximal commuting classes, there do not exist two more Pauli classes
$\cC'_{3}$ and $\cC'_{4}$ such that the common eigenbases of
$\cC_{1}$, $\cC_{2}$, $\cC_{3}'$, and $\cC_{4}'$, constitute a weakly unextendible set of four MUBs in $d=4$.
\end{theorem}
{\bf Proof:}
Let $\cC_{3}$, $\cC_{4}$, and $\cC_{5}$ denote the remaining three
classes of the complete set which are uniquely determined once
$\cC_{1}$ and $\cC_{2}$ are given.  Suppose there exist $\cC'_{3}$ and
$\cC'_{4}$ as described above. Then, $\cC'_{3}$ has to be constructed
from the elements of $\cC_{3}$, $\cC_{4}$, and $\cC_{5}$.
Theorem~\ref{thm:3max} implies that one can construct exactly one more
maximal commuting class using the elements of $\cC_{3}$, $\cC_{4}$, and
$\cC_{5}$. Let us denote this class by $\cC_{\text{unext}}$.  Thus,
$\cC'_{3}$ is either the same as $\cC_{\text{unext}}$, or it has to be
one of $\cC_{3}$, $\cC_{4}$, or $\cC_{5}$.  In the former case, there
cannot exist a $\cC'_{4}$ such that its common eigenbasis is unbiased
with respect to the other three.  In the latter case, we simply recover
a complete set of five MUBs, thus showing that we cannot obtain a
weakly unextendible set of four MUBs starting from two classes. \square

Note that Theorem~\ref{thm:d_extendible} is a special case
of~\cite{weiner09}, where it is shown that any set of $d$ MUBs in
dimension $d$ can always be extended to a complete set.

\subsection{Weakly Unextendible Sets of Five MUBs in $d=8$}\label{sec:unext_d8}

We next demonstrate a construction of weakly unextendible sets of MUBs in $d=8$. The basic construction idea is similar to that in $d=4$, but proving that such a construction always leads to an unextendible set turns out to be more complex in this case.

\begin{theorem}[{\bf Five Weakly Unextendible MUBs in $d=8$}]
Given five maximal commuting Pauli classes $\cC_{1}$,\ldots, $\cC_{5}$ that belong to a complete set of classes in dimension $d=8$, there exists exactly one more maximal commuting class that can be constructed using the elements of
$\cC_{1}\cup\ldots\cup\cC_{5}$. Denoting this new class as $\cS$,  $\{\cC_{6}, \cC_{7}, \cC_{8}, \cC_{9}, \cS\}$ is a set of five
unextendible Pauli classes, the common eigenbases of which form a set of weakly unextendible MUBs in $d=8$. \label{thm:5unext}
\end{theorem}
We know from Lemma~\ref{lem:(n-1)commute} that the only way to form a
set $\cS$ of seven commuting operators out of
$\cC_{1}\cup\ldots\cup\cC_{5}$ is to pick three elements from one class
(say $\cC_{1}$) and one element each from the remaining four
classes.  Furthermore, the three elements belonging to $\cC_{1}$ must
be of the form $\{U_{i}, U_{j}, U_{i}U_{j}\}\subset\cC_{1}$. We refer to
Appendix~\ref{sec:proof_d8} for a proof of Theorem~\ref{thm:5unext}.

The following theorem shows that starting with $k\neq 5$ classes out of a complete
set, no other maximal commuting class can be formed using the operators in $k$ such classes.
\begin{theorem}\label{thm:neq5_unext}
Given $k$ maximal commuting Pauli classes $\cC_{1}$, $\cC_{2}$,\ldots,
$\cC_{k}$ in dimension $d=8$, it is not possible to construct
another maximal commuting class using the elements of $\cC_{1}\cup
\ldots\cup\cC_{k}$ for $k\neq5$.
\end{theorem}
Once again we refer to Section~\ref{sec:proof_d8} for the
proof. Together, Theorems~\ref{thm:5unext} and ~\ref{thm:neq5_unext}
imply that it is possible to construct a set of exactly {\it five}
weakly unextendible MUBs using tensor products of Pauli operators in
$d=8$; no more, no fewer.

\subsection{Unextendible Sets in Higher Dimensions}\label{sec:unext_higher}

The existence of unextendible sets of Pauli classes in $d=4,8$ relies
entirely on properties (P1) through (P5) listed above, in particular,
Lemma~\ref{lem:(n-1)commute}. Since these properties hold for all
dimensions $d=2^{n}$, our construction of unextendible sets of classes
should generalize to higher dimensions $ d=2^{n}$ $(n\geq 3)$ as
well. However, we do not have a proof of such a general
construction yet, so we will merely conjecture the existence of
unextendible sets of $d/2+1$ MUBs here.

\begin{conjecture}[{\bf $d/2+1$ Unextendible MUBs in $d=2^{n}$}]\label{conj:unext_gen}
Given $d/2 +1$ maximal commuting Pauli classes $\cC_{1}$,
$\cC_{2}$,\ldots, $\cC_{\frac{d}{2}+1}$ that belong to a complete set
of maximal commuting classes in $d=2^{n}$, there is exactly one more
maximal commuting Pauli class $\cS$ that can be formed using the operators
of $\cC_{1}\cup\ldots\cup\cC_{\frac{d}{2}+1}$. The set of classes
$\{\cS, \cC_{\frac{d}{2}+2}, \cC_{\frac{d}{2}+3}, \ldots, \cC_{d+1}\}$
is an unextendible set of $d/2+1$ Pauli classes whose common
eigenbases form an unextendible set of MUBs.
\end{conjecture}

\section{Strongly Unextendible Sets of MUBs in $d=4,8$}\label{sec:strong_unext}

In the following we present examples of sets of three and five MUBs in $d=4$ and
$d=8$ respectively, obtained from our construction, that are in fact strongly unextendible.

Consider the following two unextendible sets of Pauli operators in $d=4$ and $d=8$.
\begin{alignat}{5}
\cC^{(4)}_{1} &= \{ Y\otimes Y, I\otimes Y,  Y\otimes I \}, \nonumber \\
\cC^{(4)}_{2} &= \{Y\otimes Z, X\otimes X,  Z\otimes Y \}, \nonumber \\
\cC^{(4)}_{3} &= \{Z\otimes I,  I\otimes Z,  Z\otimes Z \}\label{eq:pauli3_strong}
\end{alignat}
and
\begin{alignat}{5}
\cC^{(8)}_{1} &= \{IIY, YYI, YYY, IYY, YII, IYI, YIY\}, \nonumber \\
\cC^{(8)}_{2} &= \{IXI, XIX, XXX, IXX, IIX, XII, XXI\}, \nonumber \\
\cC^{(8)}_{3} &= \{ZII, IZZ, ZZZ, IIZ, IZI, ZIZ, ZZI\}, \nonumber \\
\cC^{(8)}_{4} &= \{IZX, YZI, YIX, ZYY, XYZ, ZXZ, XXY\}, \nonumber \\
\cC^{(8)}_{5} &= \{XIZ, XYI, IYZ, ZXX, YZX, YXY, ZZY\} \label{eq:pauli_d8}
\end{alignat}

Suppose there exists a normalized vector $|\psi\rangle$ that is
unbiased with respect to all joint eigenvectors of the given
classes. Since one of the eigenbases in both sets is the computational
basis we can assume that $|\psi\rangle$ is of the form
\begin{equation}\label{eq:generic_vector}
|\psi\rangle=\frac{1}{\sqrt{d}}(1,x_1,\ldots,x_{d-1})^T.
\end{equation}
Denoting the joint eigenbasis of the class $\cC_{i}$ by
$\cB_{i}=\{|b_i^{(\alpha)}\rangle\colon\alpha=1,\ldots,d\}$, we get
the following conditions on the vector $|\psi\rangle$:
\begin{equation}\label{eq:MUB_conditions}
|\langle \psi|b_i^{(\alpha)}\rangle|^2-\frac{1}{d}=0.
\end{equation}
Note that~\eqref{eq:MUB_conditions} involves complex conjugation of
the coefficients $x_j$ of the vector $|\psi\rangle$.

Unbiasedness with respect to the computational basis implies that the
coefficients $x_j$ in \eqref{eq:generic_vector} must have modulus one,
which implies that $\overline{x}_j=1/x_j$, where $\overline{x}_j$
denotes complex conjugation.  Hence the left-hand side
of~\eqref{eq:MUB_conditions} is a rational function in the $d-1$
variables $x_j$.  Equivalently, we can consider the system of
polynomial equations obtained from the numerators of the left-hand
side of~\eqref{eq:MUB_conditions}, provided that the denominator does
not vanish.  It turns out that the denominators are just products of
the variables $x_j$, so the additional condition requires that none of
the variables $x_j$ vanishes.

Using the computer algebra system Magma~\cite{Magma}, we can compute a
Gr\"obner basis for the ideal generated by the numerators of the
conditions~\eqref{eq:MUB_conditions}. From the Gr\"obner basis we can
deduce that for both the sets in~\eqref{eq:pauli3_strong} ($d=4$)
and~\eqref{eq:pauli_d8} ($d=8$), at least two of the coefficients
$x_j$ must vanish, contradicting the assumption that $|x_j|=1$. Hence,
there does not exist a vector $|\psi\rangle$ that is unbiased to all
bases, and therefore the sets of three and five MUBS
in~\eqref{eq:pauli3_strong} ($d=4$) and~\eqref{eq:pauli_d8} ($d=8$),
respectively, are strongly unextendible.

We conclude this section by explicitly writing down the strongly unextendible set of bases corresponding to the classes in~\eqref{eq:pauli3_strong}:
\begin{alignat}{5}
 \cB_{1} &= \left\{\frac{1}{2}\left(\begin{array}{c} 1 \\ i\\ i \\ -1 \end{array}\right), \frac{1}{2} \left(\begin{array}{c} 1\\ -i\\ -i\\ -1 \end{array}\right), \frac{1}{2} \left(\begin{array}{c} 1 \\ -i\\ i\\ 1 \end{array}\right), \frac{1}{2}\left(\begin{array}{c} 1\\ i\\ -i\\ 1 \end{array}\right)   \right\}                                                                                                                                                            \nonumber\displaybreak[3] \\
\cB_{2} &= \left\{ \frac{1}{2}\left(\begin{array}{c} 1 \\ -i\\ -i \\ 1 \end{array}\right), \frac{1}{2} \left(\begin{array}{c} 1\\ -i\\ i\\ -1 \end{array}\right), \frac{1}{2} \left(\begin{array}{c} 1 \\ i\\ -i\\ -1 \end{array}\right), \frac{1}{2}\left(\begin{array}{c} 1\\ i\\ i\\ 1 \end{array}\right)  \right\}                                                                                                                                                       \nonumber\displaybreak[3] \\
\cB_{3} &= \left\{ \left(\begin{array}{c} 1 \\ 0\\ 0 \\ 0 \end{array}\right), \left(\begin{array}{c} 0\\ 1\\ 0\\ 0 \end{array}\right), \left(\begin{array}{c} 0 \\ 0\\ 1\\ 0 \end{array}\right), \left(\begin{array}{c} 0\\ 0\\ 0\\ 1 \end{array}\right) \right\}
\end{alignat}

\section{Applications of Unextendible Sets in $d=4$}\label{sec:props}

Our construction of unextendible sets of classes in dimensions where a
complete set of such classes exist, offers new insight into the
structure of MUBs in these dimensions. The complete set of MUBs in
dimensions $d=2^{n}$ has $d+1$ bases, which are optimal for state
tomography, whereas the unextendible sets we construct contain
$\frac{d}{2}+1$ bases. We now discuss potential applications of such
smaller sets of MUBs for quantum foundations and for cryptographic
tasks.

\subsection{State-independent Proofs of the Kochen-Specker Theorem}

Consider the set of three Pauli classes in $d=4$ in~\eqref{eq:pauli3}. There
exists an alternate partitioning of the nine operators that constitute
the set, leading to another set of three commuting classes, namely,
\begin{alignat}{5}
\cC_{1}' &= \{Y\otimes Y, Z\otimes X, X\otimes Z\}, \nonumber \\
\cC_{2}' &= \{I\otimes Y, X\otimes Y, X\otimes I\}, \nonumber \\
\cC_{3}' &= \{Y\otimes I, Y\otimes Z, I\otimes Z\}. \label{eq:pauli3v2}
\end{alignat}
The new classes $\cC_{i}'$ are formed by picking one commuting element
each from each of $\cC_{1}$, $\cC_{2}$, and $\cC_{3}$. Each of the
nine Pauli operators in~\eqref{eq:pauli3} is a part of two maximal
commuting classes -- $\cC_{i}$ and $\cC_{j}'$. The partitions
in~\eqref{eq:pauli3} and~\eqref{eq:pauli3v2} provide two separate
contexts for each of these nine Pauli operators, thus leading to a
state-independent proof of the Kochen-Specker (KS) Theorem~\cite{KS67}
in $d=4$, similar to the proof by Mermin~\cite{mermin90}. We note that
this example of a state-independent proof of the KS Theorem is one of
nine other Mermin-like proofs obtained in~\cite{cabello_PRA10} via an
alternate approach. This set of ten Mermin-like proofs were together
shown to yield a stronger violation of
non-contextuality~\cite{cabello_PRA10,cabello_njp11} than Mermin's
proof.

The existence of two such contexts for the same set of nine operators
is a property unique to unextendible sets of classes in $d=4$, as we
will prove below. The existence of two such partitions of the same set
of nine operators is not possible for an arbitrary triple of commuting
classes that we may pick out of the complete set of five classes that
exist in $d=4$. Thus, the set of nine two-qubit Pauli operators used
in Mermin's original proof also give rise to a weakly unextendible set
of three MUBs, via a partitioning into an unextendible set of classes.

\begin{theorem}
Given an unextendible set of three maximal commuting Pauli classes
$\{\cC_{1}, \cC_{2},$ $\cC_{3}\}$, the nine operators that constitute
these classes can be partitioned into a different set of three maximal
commuting classes $\{\cC_{1}', \cC_{2}', \cC_{3}'\}$ such that
$\cC_{i}'$ has one operator each from each of $\cC_{1}$, $\cC_{2}$,
and $\cC_{3}$.
\end{theorem}
{\bf Proof:} Let us denote the unextendible set of three classes as
\begin{alignat}{5}
\cC_{1} &= \{U_{1}, V_{1}, U_{1}V_{1}\}, \nonumber \\
\cC_{2} &= \{U_{2}, V_{2}, U_{2}V_{2}\}, \nonumber \\
\cC_{3} &= \{U_{3}, V_{3}, U_{3}V_{3}\}.
\end{alignat}
If this set of three classes was in fact extendible, the operators in
$\cC_{1}$ and $\cC_{2}$ must be distributed in such a way as to
generate three more maximal commuting classes. But since these are
unextendible classes, the products of the operators in $\cC_{1}$ and
$\cC_{2}$ are distributed such that they form only one maximal
commuting class, namely, $\cC_{3}$. We have already encountered such a
distribution in proving Theorem~\ref{thm:3max}.

Let $[U_{1}, U_{2}] = 0 = [V_{1}, V_{2}]$, so that
$[U_{1}V_{1},U_{2}V_{2}] = 0$. Suppose we set $U_{3} = U_{1}U_{2}$ and
$V_{3}=V_{1}V_{2}$, so that $U_{3}V_{3} = U_{1}U_{2}V_{1}V_{2}$. Then,
as proved in Theorem~\ref{thm:3max}, the remaining product operators
$U_{1}V_{2}$, $U_{2}V_{1}$, $U_{1}U_{2}V_{2}$, $U_{2}U_{1}V_{1}$,
$V_{1}U_{2}V_{2}$, and $V_{2}U_{1}V_{1}$ cannot be used to form a maximal
commuting class. Any other assignment of commuting products to $U_{3}$
and $V_{3}$ will lead to a complete set of maximal commuting
classes. Thus the class $\cC_{3}$ for an unextendible set is of the
form
\begin{equation}
\cC_{3} = \{U_{1}U_{2}, V_{1}V_{2}, U_{1}U_{2}V_{1}V_{2}\}.
\end{equation}
Therefore, there exist three other maximal commuting classes that can be
formed using the operators in $\cC_{1}$, $\cC_{2}$, and $\cC_{3}$:
\begin{alignat}{5}
\cC_{1}' &= \{U_{1}, U_{2}, U_{1}U_{2}\}, \nonumber \\
\cC_{2}' &= \{V_{1}, V_{2}, V_{1}V_{2}\}, \nonumber \\
\cC_{3}' &= \{U_{1}V_{1}, U_{2}V_{2}, U_{1}U_{2}V_{1}V_{2}\}.
\end{alignat}
   \qquad\hfill \square

This connection with proofs of the KS Theorem is not so
straightforward for unextendible sets in $d=8$. For example, consider

the unextendible set of classes in~\eqref{eq:pauli_d8}. The operators
constituting this set do have the property that they can be
partitioned into another set of five Pauli classes, as follows:
\begin{alignat}{5}
\cC_{1}' &= \{\underbrace{IIY, YYI, YYY}_{\cC_{1}}, XXI, ZZI, XXY, ZZY\}, \nonumber \\
\cC_{2}' &= \{\underbrace{IXI, XIX, XXX}_{\cC_{2}}, ZIZ, ZXZ, YXY, YIY\}, \nonumber \\
\cC_{3}' &= \{\underbrace{ZII, IZZ, ZZZ}_{\cC_{3}}, ZYY, ZXX,IYY, IXX \}, \nonumber \\
\cC_{4}' &= \{\underbrace{IZX, YZI, YIX}_{\cC_{4}}, YZX, YII, IIX, IZI\}, \nonumber \\
\cC_{5}' &= \{\underbrace{XIZ, XYI, IYZ}_{\cC_{5}}, IYI, XII, IIZ, XYZ\}. \label{eq:pauli_d8v2}
\end{alignat}
These new classes $\cC_{i}'$ are obtained by picking three commuting
operators from the corresponding class $\cC_{i}$
in~\eqref{eq:pauli_d8} and one operator each from the remaining four
classes $\cC_{j}$ $(j\neq i)$. Thus, we have a set of operators in
$d=8$ such that every operator is part of two different maximal
commuting classes. However, unlike in the two-qubit case, this is not
sufficient to obtain a state-independent proof of the KS
Theorem.  Whether this property of the existence of two contexts for
each operator in the set holds in general for all unextendible sets in
$d=8$, and what role such sets play in proving violations of
non-contextuality, remains to be seen.

\subsection{Tightness of $H_{2}$ Entropic Uncertainty Relation}\label{sec:eur}

MUBs correspond to measurement bases that are most ``incompatible",
where the degree of incompatibility is quantified by entropic
uncertainty relations. An entropic uncertainty relation (EUR) for a
set of $L$ measurement bases $\{\cB_{1}, \cB_{2},\ldots,\cB_{L}\}$
provides a lower bound on the average entropy $H(B_{j}||\psi\rangle)$:
\begin{align}\label{eq:H1general}
\frac{1}{L} \sum_{j=1}^{L} H(B_j||\psi\rangle) \geq c_L\ ,
\end{align}
for all states $|\psi\rangle$. Here, $H(B_{j}||\psi\rangle)$ denotes
the entropy of the distribution obtained by measuring state
$|\psi\rangle \in \cS(\mathbb{C}^{d})$ in the measurement basis
$\cB_{i}$. Here we will focus on the {\it collision entropy} $H_{2}$
of the distribution obtained by measuring state $|\psi\rangle$ in the
measurement basis $\cB_{i} = \{|b_{i}^{(j)}\rangle\colon
j=1,\ldots,d\}$, defined as,
\begin{equation}
 H_{2}(\cB_{i}\vert |\psi\rangle) = -\log\sum_{j=1}^{d}(\vert\langle b_{i}^{(j)}\vert\psi\rangle\vert^{2})^{2}.
\end{equation}

For $L$ MUBs in $d$ dimensions, the collision entropy satisfies the following uncertainty relation~\cite{WWsurvey}:
\begin{equation}\label{eq:H2eur}
\frac{1}{L}\sum_{i=1}^{L}H_{2}(\cB_{i}||\psi\rangle) \geq
\log_2\left(\frac{L+d-1}{dL}\right).
\end{equation}
However, it is not known if this EUR is tight in general. Here, we use
the result of Theorem~\ref{thm:3max} to show that this uncertainty
relation is in fact tight for any three MUBs in $d=4$, whether they be
(a) part of a complete set of MUBs, or (b) a set of weakly
unextendible MUBs. We merely state the result here and refer to
Appendix~\ref{sec:EURd4} for the proof.

\begin{theorem}
Given a set of three maximal commuting Pauli classes $\{\cC_{1}, \cC_{2}, \cC_{3}\}$ in dimension $d=4$ such that at least one more maximal commuting Pauli class $\cS$ can be constructed by picking one element from each of $\cC_{1}$, $\cC_{2}$, and $\cC_{3}$. Then, the common eigenstates of the operators in $\cS$ saturate the following uncertainty relation:
\begin{equation}
 \frac{1}{3}\sum_{i=1}^{3}H_{2}(\cB_{i}||\psi\rangle) \geq  1,
\end{equation}
where $\cB_{i}$ is the common eigenbasis of the operators in $\cC_{i}$.
\end{theorem}
The tightness of the uncertainty relation in~\eqref{eq:H2eur} was
first noted in~\cite{MWB10} for the specific set of three MUBs
in~\eqref{eq:pauli3}. Here we prove that the EUR is in fact saturated
by {\it all} sets for three MUBs in $d=4$. However, unlike in the case
of $d=4$, our construction of unextendible sets in $d=8$ offers no
immediate insight into the tightness of the uncertainty relation in
$d=8$.

\section{Conclusions and Open Questions}
In this paper we have explored the question of whether there exist
smaller, unextendible sets of mutually unbiased bases in dimensions
$d=2^{n}$. We have shown by explicit construction the existence of
sets of $d/2+1$ MUBs in dimensions $d=4,8$ from Pauli classes, that
are unextendible using common eigenbases of operator classes from the
Pauli group. Our construction is based on grouping the $n$-qubit Pauli
operators into unextendible sets of $d/2+1$ maximal commuting
classes. We have shown that specific examples of such unextendible
Pauli classes in fact lead to strongly unextendible MUBs.

Since our construction relies on general properties of a complete set
of Pauli classes which hold for any $d=2^{n}$, we are led to
conjecture the existence of such unextendible classes in higher
dimensions ($n>3$) as well. Furthermore, since our construction
essentially relies on partitioning a unitary operator basis into
classes of commuting operators, it has the potential to be generalized
to the case of prime-power dimensions.

In the case of two-qubit systems we have pointed out an interesting
connection between unextendible sets of Pauli classes and
state-independent proofs of the Kochen-Specker Theorem. We have also
shown that the tightness of the $H_{2}$ entropic uncertainty relation
for any set of three MUBs in $d=4$ follows as an important consequence
of our construction.

We strongly suspect that the sets of weakly unextendible MUBs arising
from our general construction in $d=4,8$ are in fact strongly
unextendible. While we were able to prove this for specific examples
presented in this paper, a general proof remains elusive. Furthermore,
while we conjecture the existence of unextendible sets of MUBs in
dimensions $d=2^{n}$, proving this remains an open problem.

Recent works have shown that there exists a formal correspondence
between unextendible sets of MUBs and maximal partial spreads of the
polar space formed by the $n$-qubit Pauli operators in
$d=2^{n}$~\cite{thas09}. It will be interesting to study the
implications of this connection for our conjecture on the existence of
unextendible sets in higher dimensions.


\nonumsection{Acknowledgments}
SB and WKW would like to thank The Institute of Mathematical Sciences
(IMSc), Chennai, for supporting their visit during April 2012 when this work
was initiated. SB would also like to thank IMSc for supporting subsequent
visits related to this work and CQT, NUS, Singapore for supporting his
visit in November 2012.


%

\nonumsection{References}
\noindent


\appendix{: Proof of Theorem \ref{thm:3max}}
\setcounter{appendixc}{0}
\refstepcounter{appendixc}
\label{sec:proof_d4a}

Suppose we are given three mutually disjoint classes  of the form
\begin{alignat}{5}
\cS_{1} &= \{U_{1}, V_{1}, U_{1}V_{1}\}, \nonumber \\
\cS_{2} &= \{U_{2}, V_{2}, U_{2}V_{2}\}, \nonumber \\
\cS_{3} &= \{U_{3}, V_{3}, U_{3}V_{3}\}.  
\end{alignat}
Each element is $\cS_{1}$ commutes with one element from each of
$\cS_{2}$ and $\cS_{3}$. Assume without loss of generality that
$[U_{1},U_{2}]= 0 = [U_{1}, U_{3}]$ and $[V_{1},V_{2}] = 0 =
[V_{1},V_{3}]$. Note that $V_{1}$ and $U_{1}$ cannot both
  commute with the same element in $\cS_{2}$ (or $\cS_{3}$), for this
  would give a set of four commuting operators (e.\,g., $\{U_{1}, V_{1},
  U_{2}, U_{1}V_{1}\}$), which is not possible. This immediately implies that
$[U_{1}V_{1},U_{2}V_{2}]=0$ and $[U_{1}V_{1}, U_{3}V_{3}]=0$.

\noindent{\it Uniqueness:} We first show that it is not possible to
construct more than one maximal commuting set from the operators in
$\cS_{1}\cup\cS_{2}\cup\cS_{3}$. Suppose $U_{3} = U_{1}U_{2}$ and
$V_{3} = V_{1}V_{2}$, so that there exist two maximal commuting sets:
$\{U_{1}, U_{2}, U_{1}U_{2}\}$ and $\{V_{1}, V_{2}, V_{1}V_{2}\}$. The
class $\cS_{3}$ becomes
 \begin{equation}
 \cS_{3} = \{U_{1}U_{2}, V_{1}V_{2}, U_{1}U_{2}V_{1}V_{2}\}.
 \end{equation}
Now, since both $[U_{2}, U_{3}] =0$ and $[V_{2},V_{3}]=0$, we
see that $U_{2}V_{2}$ must commute with $U_{3}V_{3} =
U_{1}U_{2}V_{1}V_{2}$. Thus, the assumption that there exists more
than one maximal commuting set implies that there exists a third
maximal commuting set as well --- $\{U_{1}V_{1}, U_{2}V_{2},
U_{3}V_{3}\}$. We will show that this in fact leads to a
contradiction.

Similar to $\cS_{3}$, we can also obtain the remaining two classes
$\cS_{4}$ and $\cS_{5}$ as products of $U_{1}$, $U_{2}$, $V_{1}$,
$V_{2}$, $U_{1}V_{1}$, $U_{2}V_{2}$. Consider $U_{1}V_{2}$ and
$U_{2}V_{1}$. Note that
\begin{alignat*}{5}
[U_{1}V_{2}, U_{2}V_{1}] &= 0,\\
\text{and}\qquad(U_{1}V_{2})(U_{2}V_{1}) &= U_{1}U_{2}V_{1}V_{2} = U_{3}V_{3}.
\end{alignat*}
If $U_{1}V_{2}, U_{2}V_{1}$ were to belong to the same class, the
operator $U_{3}V_{3}$ would occur in two different classes. Therefore,
let $U_{1}V_{2} \in \cS_{4}$ and $U_{2}V_{1} \in \cS_{5}$. Next,
consider the products $U_{1}(U_{2}V_{2})$ and
$U_{2}(U_{1}V_{1})$. Since $U_{1}(U_{2}V_{2}) = (U_{1}V_{2})V_{2}$,
and we have assumed $U_{1}V_{2} \in \cS_{4}$, this operator cannot
belong to $\cS_{4}$. And, since
\begin{alignat}{5}
&&(U_{2}V_{1}) (U_{1}U_{2}V_{2}) &= - U_{1}V_{1}V_{2}, \nonumber \\
&&(U_{1}U_{2}V_{2})(U_{2}V_{1}) &= U_{1}V_{1}V_{2}, \nonumber \\
\Rightarrow&& [U_{2}V_{1}, U_{1}U_{2}V_{2}] &\neq 0,
\end{alignat}
$U_{1}(U_{2}V_{2})$ cannot belong to $\cS_{5}$ either. Similarly, the
product $U_{2}(U_{1}V_{1})$ cannot belong to $\cS_{4}$ or
$\cS_{5}$. Thus, our assumption that there exist two maximal commuting
classes in $\cS_{1}\cup \cS_{2}\cup \cS_{3}$ leads to a situation where
there are not enough commuting operators to form the remaining two
classes $\cS_{4}$ and $\cS_{5}$.

On the other hand, suppose we had assumed $U_{3} = U_{1}V_{2}$ and
$V_{3} = U_{2}V_{1}$, we would have exactly one maximal commuting
class, namely,
\begin{equation}
\{U_{1}V_{1}, U_{2}V_{2}, U_{3}V_{3} = U_{1}U_{2}V_{1}V_{2}\}
\end{equation}
and a sufficient number of commuting operators to form the remaining two classes:
\begin{eqnarray}
\cS_{4} &=& \{U_{1}U_{2}, V_{1}U_{2}V_{2}, V_{2}U_{1}V_{1}\}, \nonumber \\
\cS_{5} &=& \{V_{1}V_{2}, U_{1}U_{2}V_{2}, U_{2}U_{1}V_{1}\}. \nonumber
\end{eqnarray}
\noindent{\it Existence:} We next show that there has to exist at
least one maximal commuting class (distinct from $\cS_{1}$, $\cS_{2}$,
$\cS_{3}$) that can be formed using the set
$\{U_{i},V_{i},U_{i}V_{i}\}_{i=1}^{3}$. Once again, let $\cS_{4}$ and
$\cS_{5}$ denote the two remaining classes that form a complete
set. Consider the elements $U_{1}V_{2}$ and $V_{1}V_{2}$. These cannot
belong to either $\cS_{1}$ or $\cS_{2}$, by construction. Suppose,
$U_{1}U_{2}, V_{1}V_{2} \notin \cS_{3}$ either. Furthermore,
$U_{1}U_{2}$ and $V_{1}V_{2}$ each have to belong to a different
class, say $\cS_{4}$ and $\cS_{5}$, respectively. For, if they belonged
to the same class, say $\cS_{4}$, we could form more than one maximal
commuting class from the operators in $\cS_1$, $\cS_2$, $\cS_4$,
contradicting the uniqueness result above.

Now, consider the operator $U_{1}V_{1}U_{2}V_{2}$. This also cannot
belong to the classes $\cS_{1}$ or $\cS_{2}$, by
construction. Further, note that,
\begin{alignat*}{5}
 [(U_{1}V_{1})(U_{2}V_{2}) , U_{1}U_{2}]
&= (U_{1}U_{2})(U_{1}V_{1})(U_{2}V_{2}) - (U_{1}V_{1})(U_{2}V_{2})(U_{1}U_{2})\\
&= U_{2}(V_{1}V_{2})U_{2} - U_{1}(V_{1}V_{2})U_{1}
\end{alignat*}
Therefore $U_{1}V_{1}U_{2}V_{2}$ commutes with $U_{1}U_{2}$ iff
$U_{1}=U_{2}$. Similarly, we argue that it does not commute with
$V_{1}V_{2}$ either, implying that it cannot belong to the remaining
two classes $\cS_{4}$ and $\cS_{5}$. Thus, $U_{1}V_{1}U_{2}V_{2} \in
\cS_{3}$, leading to the existence of at least one maximal commuting
class: $\{U_{1}V_{1}, U_{2}V_{2}, U_{1}V_{1}U_{2}V_{2}\}$. \square

\label{sec:proof_d4}
\appendix{: Five Unextendible Classes in $d=8$}\label{sec:proof_d8}

\subappendix{Proof of Theorem~\ref{thm:5unext}}
\begin{theorem}[Five Weakly Unextendible MUBs in $d=8$]\hskip0em plus 2em
Given a set of five maximal commuting classes $\{\cC_{1},\ldots,
\cC_{5}\}$ which are taken from a complete set of classes in dimension
$d=8$, there exists exactly one more maximal commuting class that can
be constructed using the elements of
$\cC_{1}\cup\ldots\cup\cC_{5}$. Denoting this new class as $\cS$, the
set $\{\cC_{6}, \cC_{7}, \cC_{8}, \cC_{9}, \cS\}$ is a set of five
unextendible MUBs in $d=8$.
\end{theorem}
{\bf Proof:}
We know from Lemma~\ref{lem:(n-1)commute} that the only way to form a
set $\cS$ of seven commuting operators out of
$\cC_{1}\cup\ldots\cup\cC_{5}$ is to pick three elements from one
class (say $\cC_{1}$) and one element each from the remaining four
classes. Furthermore, the three elements belonging to $\cC_{1}$ must
be of the form $\{U_{i}, U_{j}, U_{i}U_{j}\}\subset\cC_{1}$.

{\it Existence:} We first show that such a maximal commuting class
$\cS$ can always be found, given any five maximal commuting classes.
Note that there are seven such distinct triples that can formed using
the elements of $\cC_{1}$. Once we pick three operators in $\cC_{1}$,
there is a unique element $V_{(ij)} \in \cC_{2}$ that commutes with
the first three. The remaining operators in $\cS$ will therefore have
to be $U_{i}V_{(ij)}$, $U_{j}V_{(ij)}$, and $U_{i}U_{j}V_{(ij)}$. Our
task is to show that at least one of the seven triples of the form
$\{U_{i},U_{j},U_{i}U_{j}\}$ is such that the corresponding operators
$U_{i}V_{(ij)}$, $U_{j}V_{(ij)}$, and $U_{i}U_{j}V_{(ij)}$ must belong to
the classes $\cC_{3}$, $\cC_{4}$, and $\cC_{5}$ respectively. This
follows once we make the following observations:
\begin{itemize}
\item [(T1)] The three operators $U_{i}V_{(ij)}$, $U_{j}V_{(ij)}$, and
  $U_{i}U_{j}V_{(ij)}$ should each belong to a different class.  Clearly,
  all three cannot belong to the same class, for that would imply that
  $U_{i}U_{j}$ occurs in two different classes.  No two of them can
  belong to the same class either, for the third is simply a product
  of the other two.
\item [(T2)] Let us label the seven triples that can be formed using the
  elements of $\cC_{1}$ as $\tau_{1}$, $\tau_{2}$,\ldots, $\tau_{7}$.
  Any two of them share exactly one common element, that is,
  $|\tau_{i}\cap\tau_{j}| = 1$ for $i\ne j$. Consider two such triples
  of the form $\tau_{1} = \{U_{i}, U_{j}, U_{i}U_{j}\}$ and $\tau_{2}
  = \{U_{i}, U_{k}, U_{i}U_{k}\}$. Say $V_{(ij)} \in \cC_{2}$ is the
  unique operator in $\cC_{2}$ that commutes with $\tau_{1}$, and
  $V_{(ik)}\in \cC_{2}$ the operator that commutes with
  $\tau_{2}$. The triples obtained by multiplying with the
  corresponding commuting elements in $\cC_{2}$ are distributed among
  the remaining classes such that the operators $U_{i}V_{(ij)}$ and
  $U_{i}V_{(ik)}$ belong to different classes. Thus, the elements of
  any two triples $\tau_{i}$ and $\tau_{j}$ cannot be distributed
  within three classes, but need four classes.
\end{itemize}
In order to satisfy the above constraints, the operators obtained
as products of the triples $\tau_{1}$, $\tau_{2}$, \ldots, $\tau_{7}$ with
the corresponding commuting operators in $\cC_{2}$, must be
distributed as follows:
\begin{equation}\label{eq:triples}
\begin{array}[b]{cccccccccc}
\tau_{1} &\rightarrow&        &      &      &\cC_{6}&\cC_{7}&\cC_{8}\\
\tau_{2} &\rightarrow&        &      &      &      &\cC_{7}&\cC_{8}&\cC_{9}\\
\tau_{3} &\rightarrow&\cC_{3}&       &      &      &      &\cC_{8}&\cC_{9}\\
\tau_{4} &\rightarrow&\cC_{3}& \cC_{4}&\cC_{5}&\\
\tau_{5} &\rightarrow&        &\cC_{4}&\cC_{5}&\cC_{6}\\
\tau_{6} &\rightarrow&        &      &\cC_{5}&\cC_{6}&\cC_{7}\\
\tau_{7} &\rightarrow&        &      &      &\cC_{6}&\cC_{7}&\cC_{8}\\
\end{array}
\end{equation}
This completes our proof of the existence of at least one triple of
operators $\{U_{i}, U_{j}, U_{i}U_{j}\} \in \cC_{1}$ and the
corresponding commuting operator $V_{(ij)} \in \cC_{2}$, such that
$U{i}V_{(ij)} \in \cC_{3}$, $U_{j}V_{(ij)} \in \cC_{4}$, and
$U_{i}U_{j}V_{(ij)} \in \cC_{5}$. Thus, we have at least one maximal
commuting class $\cS$ as desired.

{\it Uniqueness:} The distribution of triples in~\eqref{eq:triples}
shows that there exists exactly one triple of the $\{U_{i}, U_{j},
U_{i}U_{j}\}\subset\cC_{1}$ which can lead to a maximal commuting
class $\cS$ comprising of elements from
$\cC_{1}\cup\cC_{2}\cup\ldots\cup\cC_{5}$. The question remains as to
whether we can find another maximal commuting class $\cS'$ starting
with two generators from a class other than $\cC_{1}$.  We will now
argue that this is impossible using the fact that $\{\cC_{1}, \ldots,
\cC_{5}\}$ is extendible to a complete set of nine maximal commuting
classes.

We begin by noting that the complete set can be generated starting
with any two maximal commuting classes, say $\cC_{1}$ and
$\cC_{2}$.  Let $\{A_{1}, A_{2}, A_{3}\}$ denote a set of generators
for $\cC_{1}$ and $\{B_{1}, B_{2}, B_{3}\}$ be a set of generators for
$\cC_{2}$.  Each generator of $\cC_{2}$ commutes with at most two
generators of $\cC_{1}$ and vice-versa.  Without loss of generality,
let us assume
\begin{alignat*}{5}
\left[A_{1}, B_{3} \right] = &0 = \left[ A_{2},B_{3} \right], \\
\left[A_{2}, B_{1} \right] = &0 = \left[ A_{3},B_{1} \right], \\
\left[A_{3}, B_{2} \right] = &0 = \left[ A_{1}, B_{2} \right].
\end{alignat*}

Then, the generators of all classes can be denoted as follows:
\begin{alignat*}{5}
\cC_{1} &\equiv \langle A_{1}, A_{2}, A_{3} \rangle \\
\cC_{2} &\equiv \langle B_{1}, B_{2}, B_{3} \rangle \\
\cC_{3} &\equiv \langle A_{1}B_{1}, A_{2}B_{2}, A_{3}B_{3} \rangle \\
\cC_{4} &\equiv \langle A_{1}B_{3}, A_{2}(B_{2}B_{3}), A_{3}(B_{1}B_{2})\rangle \\
\cC_{5} &\equiv \langle A_{1}B_{2}, A_{3}(B_{2}B_{3}), A_{2}(B_{1}B_{2}B_{3})\rangle \\
\cC_{6} &\equiv \langle A_{2}B_{1}, A_{1}(B_{2}B_{3}), A_{3}(B_{1}B_{3}) \rangle \\
\cC_{7} &\equiv \langle A_{2}B_{3}, A_{1}(B_{1}B_{3}), A_{3}(B_{1}B_{2}B_{3}) \rangle \\
\cC_{8} &\equiv \langle A_{3}B_{1}, A_{2}(B_{1}B_{2}), A_{1}(B_{1}B_{2}B_{3}) \rangle \\
\cC_{9} &\equiv \langle A_{3}B_{2}, A_{1}(B_{1}B_{2}), A_{2}(B_{1}B_{3}) \rangle
\end{alignat*}

While this construction might appear rather specific, in fact any
complete set of maximal commuting classes in $d=8$ can be realized in
this fashion. In other words, given any two classes $\cC_{1}$ and
$\cC_{2}$ in $d=8$, we can always identify a set of generators $\{
A_{1}, A_{2}, A_{3} \}\subset\cC_{1}$ and $\{ B_{1}, B_{2},
B_{3}\}\subset\cC_{2}$ that generate the rest of the classes as
described above.

Within such a realization of the complete set of maximal commuting
classes, consider some set of five classes, for example, $\{\cC_{1},
\cC_{2}, \cC_{4}, \cC_{7}, \cC_{8}\}$.
\begin{alignat}{5}
\cC_{1} &\equiv \langle \underline{A_{1}}, \underline{A_{2}}, A_{3} \rangle \nonumber \\
\cC_{2} &\equiv \langle B_{1}, B_{2}, \underline{B_{3}} \rangle \nonumber \\
\cC_{4} &\equiv \langle \underline{A_{1}B_{3}}, A_{2}(B_{2}B_{3}), A_{3}(B_{1}B_{2})\rangle \nonumber \\
\cC_{7} &\equiv \langle \underline{A_{2}B_{3}}, A_{1}(B_{1}B_{3}), A_{3}(B_{1}B_{2}B_{3}) \rangle \label{eq:5classesd8} \\
\cC_{8} &\equiv \langle A_{3}B_{1}, A_{2}(B_{1}B_{2}), A_{1}(B_{1}B_{2}B_{3}) \rangle \ni \underline{(A_{1}A_{2})B_{3}} \nonumber
\end{alignat}
As proved earlier, there exists a maximal commuting class $\cS$ that
can be formed out of the operators in these five classes. Following our
construction, the class $\cS$ is obtained by choosing the generators
$A_{1}, A_{2} \in \cC_{1}$ and the commuting generator $B_{3}\in
\cC_{2}$:
\begin{equation}
 \cS = \{A_{1}, A_{2}, B_{3}, A_{1}B_{3}, A_{2}B_{3}, A_{1}A_{2}B_{3}\}. \nonumber
\end{equation}

This explicit construction allows us to see that $\cS$ is in fact the
only maximal commuting class that can be constructed from these five
classes. We have already seen that such a class cannot be formed by
choosing two other generators from $\cC_{1}$. Starting with two
generators $B_{1}, B_{2}\in \cC_{2}$ and the commuting generator
$A_{3}\in \cC_{1}$, the resulting class has at least one operator
$A_{3}B_{2}$ that is not contained in $\cC_{4}$, $\cC_{7}$, or $\cC_{8}$.

Pairs of generators in the remaining classes are of three different
types: (a) Suppose we choose $\{A_{i}(B_{i}B_{j}),
A_{j}(B_{i}B_{j}B_{k})\}$ to start with. The corresponding commuting
operator in $\cC_{1}$ is $A_{i}A_{j}$. Taking products, the resulting
class has no operator which is only a product of the $B_{i}$'s, and
therefore no operator from $\cC_{2}$. The resulting class therefore
contains at least one operator that is outside of the given five
classes. (b) Suppose we choose $\{A_{i}(B_{i}B_{j}),
A_{j}(B_{i}B_{k})\}$. The corresponding commuting operator from
$\cC_{1}$ is $A_{i}A_{j}A_{k}$, the resulting class therefore has no
operator from $\cC_{2}$ and cannot be formed using the operators in
these five classes.  (c) Choosing $\{A_{i}B_{k}, A_{j}(B_{j}B_{k})\}$,
the corresponding commuting generators are $A_{i}\in \cC_{1}$ and
$B_{k} \in \cC_{2}$. But taking products, the operator $A_{j}B_{j}$ is
not contained in this set of five classes. We have thus shown that it
is not possible to find two generators in $\cC_{2}$, $\cC_{4}$,
$\cC_{7}$, or $\cC_{8}$, such that, along with the commuting generator
from $\cC_{1}$, they generate a different maximal commuting class
within these five classes.

Finally, we note than any set of five classes can be realized as
described in Equation~\eqref{eq:5classesd8}, once the sets of generators
$\{A_{1}, A_{2}, A_{3}\}$ and $\{B_{1},B_{2},B_{3}\}$ are suitably
identified. Our uniqueness argument is therefore completely general,
and shows that there exists only one more maximal commuting class in
any set of five classes in $d=8$. \square

\subappendix{Proof of Theorem~\ref{thm:neq5_unext}}
\begin{theorem}
Given $k$ maximal commuting classes $\cC_{1}$, $\cC_{2}$,\ldots,
$\cC_{k}$ in dimension $d=8$, it is not possible to construct
another maximal commuting class using the elements of $\cC_{1}\cup
\ldots\cup\cC_{k}$ for $k\neq5$.
\end{theorem}
{\bf Proof:}
\begin{itemize}
\item[(i)]{\bf $k=3$\normalfont{:}} We first consider starting with
  $\cC_{1}$, $\cC_{2}$, and $\cC_{3}$.  Since no more than three elements
  in a given class can commute with a fixed element of a different
  class, there are only two ways to construct a maximal commuting
  class using the elements of $\cC_{1}\cup\cC_{2}\cup\cC_{3}$:
\begin{itemize}
\item[(a)] find three elements each from two of the classes (say
  $\cC_{1}$ and $\cC_{2}$) and one more element from $\cC_{3}$ that
  all commute, or
\item[(b)] find three elements from $\cC_{1}$ and two elements each from
  $\cC_{2}$ and $\cC_{3}$ that mutually commute.
\end{itemize}
But we know from Lemma~\ref{lem:(n-1)commute} that a given set of three
operators in $\cC_{1}$ cannot commute with more than one element from
either $\cC_{2}$ or $\cC_{3}$. Thus, both constructions (a) and (b)
are ruled out, and hence we cannot find a fourth maximal commuting
class, given three maximal commuting classes.

The preceding argument also rules out the case of $k=2$ classes, as
we cannot chose more than three elements form a given class.

\item[(ii)]{\bf $k=4$\normalfont{:}} Say we start with
  $\cC_{1}$,\ldots, $\cC_{4}$. In order to construct a maximal
  commuting set of seven operators form the elements of these four
  classes, we will again have to find two elements in one class that
  commute with three elements of a different class.  This is not
  possible, as shown in Lemma~\ref{lem:(n-1)commute}.  Thus, given
  four maximal commuting classes, we cannot form a fifth class using
  their elements.

\item[(iii)]{\bf $k=6$\normalfont{:}} Given six maximal commuting
  classes in $d=8$, the only way to construct another maximal
  commuting class is to pick one generator each from five of the
  classes (say $\cC_{1}$,\ldots, $\cC_{5}$), and two elements from
  $\cC_{6}$. However, Lemma~\ref{lem:(n-1)commute} implies that it is
  not possible to pick exactly {\it two} operators from a single class
  that commute with {\it one} operator in a different class, in
  $d=8$. The product of the two elements from $\cC_{6}$ gives a third
  operator in $\cC_{6}$ that also commutes with the other elements,
  thus exceeding the limit of seven commuting Pauli operators.

\item[(iv)]{$k=7$:} Given seven maximal commuting classes in $d=8$, the only way to
construct another maximal commuting class is to find one element in
each class such that all seven of them mutually commute.  Note that
given seven distinct elements, at least three of them must be
independent.  Assume that we pick three independent elements from the
first three classes, i.\,e., $A_{1}\in\cC_{1}$, $B_{1}\in\cC_{2}$, and
$C_{1}\in\cC_{3}$, Then the rest of the new maximal commuting class
$\cS$ is given by the products of these three operators, that is, we
have
\begin{equation}
\cS = \{A_{1}, B_{1}, C_{1}, A_{1}B_{1}, A_{1}C_{1}, B_{1}C_{1}, A_{1}B_{1}C_{1}\}. \nonumber
\end{equation}
Now, suppose $A_{1}B_{1} \in \cC_{4}$. According to
Lemma~\ref{lem:(n-1)commute}, corresponding to each element in $\cS$,
there exist commuting triples of the form $\{A_{1}B_{1}, D_{i},
(A_{1}B_{1})D_{i}\} \in \cC_{4}$. Since $\cC_{4}$ has only three
independent generators --- $A_{1}B_{1}$, $D_{1}$, and $D_{2}$ ---
there exist only three such triples, namely:
\begin{alignat*}{5}
\tau_{1} &= \{A_{1}B_{1}, D_{1}, (A_{1}B_{1})D_{1}\}, \\
\tau_{2} &= \{A_{1}B_{1}, D_{2}, (A_{1}B_{1})D_{2}\}, \\
\tau_{3} &= \{A_{1}B_{1}, D_{1}D_{2}, (A_{1}B_{1})(D_{1}D_{2})\}.
\end{alignat*}
Say $A_{1}$ commutes with $\tau_{1}$, $B_{1}$ commutes with $\tau_{2}$,
and $C_{1}$ commutes with $\tau_{3}$.  The operator $B_{1}C_{1}$ has to
commute with one of the triples $\tau_{1}$, $\tau_{2}$, or
$\tau_{3}$. This in turn leads to a set of four independent, commuting
generators, which cannot exist in $d=8$.  For example, if $B_{1}C_{1}$
commutes with the operators in $\tau_{1}$, $D_{1}$ commutes with
$A_{1}$, $A_{1}B_{1}$, and $B_{1}C_{1}$, leading to a set of four
independent operators, all of which commute.

We have therefore shown that it is not possible to construct a maximal
commuting class by picking one element each from $k=7$ such classes in
$d=8$.  \square
\end{itemize}

\appendix{: Tightness of $H_{2}$ Uncertainty Relation}\label{sec:EURd4}

Before we prove our result on the tightness of the entropic
uncertainty relation (EUR), we introduce a parameterization of the MUB
vectors obtained from commuting classes of Pauli operators, which will prove
useful in evaluating the $H_{2}$ entropy.

\subappendix{Parameterizing MUB Vectors in Terms of Binary Strings}

Given a set of maximal commuting classes of Pauli operators in $d=4$, let
$\sigma_{i}^{(j)}$ denote the $j$th element in the class
$\cC_{i}$.  The vectors $\{|b_{i}^{(\alpha)}\rangle \}$ of the basis
$\cB_{i}$ corresponding to the class $\cC_{i}$ can be parameterized in
terms of binary strings $\vec{\alpha} =
(\alpha_{1},\alpha_{2},\alpha_{3})$ as follows:
\begin{equation}
 |b_{i}^{(\alpha)}\rangle\langle b_{i}^{(\alpha)}| = \frac{1}{4}\left[\Id + \sum_{j=1}^{3} (-1)^{\alpha_{j}}\sigma_{i}^{(j)}\right], \; \; \alpha_{j} \in \{0,1\}.
\end{equation}
Note that the states $|b_{i}^{(\alpha)}\rangle\langle
b_{i}^{(\alpha)}|$ are guaranteed to be pure states, since they are
Hermitian and for all $i$ and $\alpha$, we have
\begin{equation}
\tr\left[|b_{i}^{(\alpha)}\rangle\langle b_{i}^{(\alpha)}|\right]^{2} = \frac{4}{16}\tr[\Id] = 1 = \tr\left[|b_{i}^{(\alpha)}\rangle\langle b_{i}^{(\alpha)}|\right].
\end{equation}
Each basis $\cB_{i} = \{|b_{i}^{(\alpha)}\rangle,
|b_{i}^{(\beta)}\rangle,
|b_{i}^{(\gamma)}\rangle,|b_{i}^{(\delta)}\rangle \}$ is thus
parameterized by four $3$-bit binary strings $\vec{\alpha}$,
$\vec{\beta}$ ,$\vec{\gamma}$, $\vec{\delta}$ which satisfy the
following property.

\begin{lemma}
The binary strings $\vec{\alpha}$, $\vec{\beta}$, $\vec{\gamma}$,
$\vec{\delta}$ that parameterize the vectors of a basis $\cB_{i} =
\{|b_{i}^{(\alpha)}\rangle, |b_{i}^{(\beta)}\rangle,
|b_{i}^{(\gamma)}\rangle,|b_{i}^{(\delta)}\rangle \}$ are such that
for any $j=1,2,3$, the string
$\alpha_{j}\beta_{j}\gamma_{j}\delta_{j}$ has Hamming weight $2$.
\end{lemma}
{\bf Proof:}
For any two vectors $\alpha\neq\beta$ in the basis $\cB_{i}$, $\langle
b_{i}^{(\beta)}|b_{i}^{(\alpha)}\rangle = \delta_{\alpha\beta}$, which implies
\begin{alignat*}{5}
\tr\left[|b_{i}^{(\alpha)}\rangle\langle  b_{i}^{(\alpha)}|b_{i}^{(\beta)}\rangle\langle  b_{i}^{(\beta)}|\right]
&{}= \frac{1}{16}\tr\left[\Id + \Id\sum_{j=1}^{3}(-1)^{\alpha_{j}\oplus\beta_{j}}\right]\\
&{}=  \frac{1}{4}\left[1 + \sum_{j=1}^{3}(-1)^{\alpha_{j}\oplus\beta_{j}}\right] = 0.
\end{alignat*}
This implies that for $\alpha\neq\beta$, the strings $\vec{\alpha}$
and $\vec{\beta}$ can coincide in only one location, i.e., there is
precisely one value of $i$ for which $\alpha_{i}\oplus\beta_{i} =
0$. A more formal statement would be that the strings $\vec{\alpha}$
and $\vec{\beta}$ have a Hamming distance of $2$. The Hamming distance between two binary strings of equal length is the number of positions at which the corresponding bits are different.

Each basis $\cB_{i} = \{|b_{i}^{(\alpha)}\rangle,
|b_{i}^{(\beta)}\rangle,
|b_{i}^{(\gamma)}\rangle,|b_{i}^{(\delta)}\rangle \}$ is thus
parameterized by four $3$-bit binary strings $\vec{\alpha}$,
$\vec{\beta}$ ,$\vec{\gamma}$, $\vec{\delta}$ which are at  Hamming
distance of $2$ from each other. Ignoring the bit where two strings
take on the same value, the remaining $2$-bit strings must be at a
Hamming distance of $2$.  Notice that for a given $2$-bit string, there
is a unique $2$-bit string that is at Hamming distance $2$ from
it. Thus, if $\alpha_{i}\oplus\beta_{i} = 0$ for some $i=1,2,3$,
$\alpha_{i}\oplus\gamma_{i} = 0$ would imply that the remaining two
bits of $\vec{\gamma}$ are the same as those of $\vec{\beta}$ making
$\vec{\gamma} = \vec{\beta}$.

For a given $i$ therefore, $\alpha_{i} \oplus\beta_{i} = 0$ would
imply that $\alpha_{i}\oplus\gamma_{i} =\alpha_{i}\oplus\delta_{i}
=1$, so that the string $\alpha_{i}\beta_{i}\gamma_{i}\delta_{i}$ has
exactly two $0$'s and two $1$'s. \square

\subappendix{Tightness of EUR}

Now we are ready to prove the tightness of the EUR for the collision
entropy for sets of three MUBs in $d=4$.

\begin{theorem}
Assume that we are given a set of three maximal commuting classes
$\{\cC_{1},$ $\cC_{2},\cC_{3}\}$ in dimension $d=4$ such that at least
one maximal commuting class $\cS$ can be constructed by picking one
element from each of $\cC_{1}$, $\cC_{2}$, and $\cC_{3}$. Then, the
common eigenstates of the operators in $\cS$ saturate the following
uncertainty relation:
\begin{equation}
 \frac{1}{3}\sum_{i=1}^{3}H_{2}(\cB_{i}||\psi\rangle) \geq  1,
\end{equation}
where $\cB_{i}$ is the common eigenbasis of the operators in
$\cC_{i}$.
\end{theorem}
{\bf Proof:}
Suppose $\{\sigma_{1}^{(1)}, \sigma_{2}^{(2)}, \sigma_{3}^{(3)}\}$
form a commuting set. Then, the common eigenstates of such a set can
again be denoted as
\begin{equation}
|\psi\rangle\langle\psi| = \frac{1}{4}\left[\Id + (-1)^{e_{1}}\sigma_{1}^{(1)} + (-1)^{e_{2}}\sigma_{2}^{(2)}+ (-1)^{e_{3}}\sigma_{3}^{(3)}\right]. \nonumber
\end{equation}

Recall that the collision entropy corresponding to measuring the state
$|\psi\rangle$ in the basis $\cB_{i} \equiv \{|b_{i}^{(x)}\rangle\colon
x=\alpha,\beta,\gamma,\delta\}$ is given by
\begin{equation}
H_{2}(\cB_{i}||\psi\rangle) = -\log\sum_{x=\alpha,\beta,\gamma,\delta}\left(\tr\bigl[|b_{i}^{(x)}\rangle\langle b_{i}^{(x)}|\psi\rangle\langle\psi|\bigr]\right)^{2}. \nonumber
\end{equation}
Expanding $|\psi\rangle$ and $\{|b_{i}^{(x)}\rangle\}$ in terms of the
operators $\sigma$ as described above, we see that for a given choice
of basis $\cB_{i}$, only those coefficients in the expansion of
$|\psi\rangle$ that correspond to operators in $\cC_{i}$ contribute to
the collision entropy. Thus, for $i=1$,
\begin{alignat*}{5}
H_{2}(\cB_{1}|\psi\rangle) &=-\log\sum_{x=\alpha,\beta,\gamma,\delta}
\left(\tr\bigl[|b_{1}^{(x)}\rangle\langle b_{1}^{(x)}|\psi\rangle\langle\psi|\bigr]\right)^{2} \\
&= -\log\sum_{x=\alpha,\beta,\gamma,\delta}\left(\frac{1}{16}\tr\bigl[\Id + (-1)^{x_{1}\oplus e_{1}}\Id\bigr]\right)^{2} \\
&= -\log\sum_{x=\alpha,\beta,\gamma,\delta}\left(\frac{4}{16}[1 + (-1)^{x_{1}\oplus e_{1}}]\right)^{2} \\
&= -\log\frac{1}{16}\sum_{x=\alpha,\beta,\gamma,\delta}\left[1 + (-1)^{x_{1}\oplus e_{1}}\right]^{2}.
\end{alignat*}
Since the string $\alpha_{1}\beta_{1}\gamma_{1}\delta_{1}$ has
exactly two $0$'s and two $1$'s, the expression $\left[1 +
  (-1)^{x_{1}\oplus e_{1}}\right]$ correspondingly takes on the values
$0$ and $2$. Since the former property holds for any string
$\alpha_{i}\beta_{i}\gamma_{i}\delta_{i}$ (for all $i\in[1,5]$), the
expression $\left[1 + (-1)^{x_{i}\oplus e_{i}}\right]$ takes on the
value $0$ half the time and the value $2$ half the time, independent
of the value of $i$. Therefore, for $i=1,\ldots 5$,
\begin{equation}
 H_{2}(\cB_{i}||\psi\rangle) = -\log\frac{1}{16}[0+0+4+4] = 1. \nonumber
\end{equation}
\qquad\hfill \square
\end{document}